\documentclass{ifacconf}
\usepackage{soul}

\usepackage{algorithm}
\usepackage{amsmath}
\usepackage{amssymb}

\usepackage{array}
\newcolumntype{L}[1]{>{\raggedright\arraybackslash}p{#1}}   
\newcolumntype{C}[1]{>{\centering\arraybackslash}p{#1}}     
\newcolumntype{R}[1]{>{\raggedleft\arraybackslash}p{#1}}    

\usepackage{graphicx}      

\usepackage{mathrsfs}
\usepackage{mathtools}

\usepackage[square,authoryear]{natbib}

\usepackage{pgfplots}
\pgfplotsset{compat=1.18}
\usepackage{subcaption}

\usepackage{multirow} 

\usepackage{natbib}        

\usepackage{hyperref}
\usepackage{acronym} 

\usepackage{tikz}
\usepackage{titlesec}
\AtBeginDocument{\setcounter{secnumdepth}{3}}
\renewcommand\thesubsubsection{\thesection.\arabic{subsection}.\roman{subsubsection}}

\titleformat{\subsubsection}[block]{\normalfont\itshape}{\thesubsubsection.}{0.5em}{}
\titlespacing*{\subsubsection}{0pt}{1ex plus .2ex}{0.8ex} 

\AtBeginDocument{}
\AtBeginDocument{}
\AtBeginDocument{}

\newtheorem{theorem}{Theorem}

\newtheorem{definition}{Definition}

\newtheorem{assumption}{Assumption}

\newcommand{\col}{\text{col}}

\newcommand{\diag}{\text{diag}}

\newcommand{\norm}[1]{\left\| #1 \right\|}

\newcommand{\bracketl}[1]{\left[ #1 \right]}

\newcommand{\quadsupply}[3]{\begin{bmatrix}
    \by \\ \bu
\end{bmatrix}^{\hspace{-1pt}T}\hspace{-4pt}\begin{bmatrix}
    #1 & #2 \\ * & #3
\end{bmatrix}\hspace{-3pt}\begin{bmatrix}
    \by \\ \bu
\end{bmatrix}}
\newcommand{\quadsupplyinput}[5]{\begin{bmatrix}
    #4 \\ #5
\end{bmatrix}^{\hspace{-1pt}T}\hspace{-4pt}\begin{bmatrix}
    #1 & #2 \\ * & #3
\end{bmatrix}\hspace{-3pt}\begin{bmatrix}
    #4 \\ #5
\end{bmatrix}}

\newcommand{\dbx}{\dot{\textbf{x}}}

\newcommand{\cL}{\mathcal{L}}

\newcommand{\bbA}{\mathbb{A}}

\newcommand{\bbN}{\mathbb{N}}

\newcommand{\bbR}{\mathbb{R}}
\newcommand{\bbS}{\mathbb{S}}

\newcommand{\bbZ}{\mathbb{Z}}

\newcommand{\bzero}{\mathbf{0}}

\newcommand{\bA}{\mathbf{A}}
\newcommand{\bB}{\mathbf{B}}

\newcommand{\bH}{\mathbf{H}}
\newcommand{\bI}{\mathbf{I}}

\newcommand{\bK}{\mathbf{K}}

\newcommand{\bP}{\mathbf{P}}
\newcommand{\bQ}{\mathbf{Q}}
\newcommand{\bR}{\mathbf{R}}
\newcommand{\bS}{\mathbf{S}}

\newcommand{\bX}{\mathbf{X}}

\newcommand{\be}{\mathbf{e}}

\newcommand{\bu}{\mathbf{u}}

\newcommand{\bx}{\mathbf{x}}
\newcommand{\by}{\mathbf{y}}

\newcommand{\scrG}{\mathscr{G}}

\newcommand{\hatbH}{\widehat{\bH}}

\newcommand{\hatbQ}{\widehat{\bQ}}
\newcommand{\hatbR}{\widehat{\bR}}
\newcommand{\hatbS}{\widehat{\bS}}

\newcommand{\tilbQ}{\widetilde{\bQ}}
\newcommand{\tilbR}{\widetilde{\bR}}
\newcommand{\tilbS}{\widetilde{\bS}}

\newcommand{\thh}{\mathrm{th}}

\acrodef{vndt}[{NDT}]{Network Dissipativity Theorem}
\acrodef{io}[{IO}]{input-output}

\acrodef{lmi}[{LMI}]{linear matrix inequality}
\acrodef{lqr}[{LQR}]{linear quadratic regulator}

\acrodef{uav}[{UAV}]{unmanned aerial vehicles}
\acrodef{cps}[{CPS}]{cyber-physical system}
\newcommand{\hatscrG}{\widehat{\mathscr{G}}}

\newcommand{\hati}{\hat{i}}

\newcommand{\numberthis}{\refstepcounter{equation}\tag{\theequation}}
\begin{document}
\sloppy
\begin{frontmatter}

\title{Dissipativity and $\cL_2$ Stability of Large-Scale Networks with Changing Interconnections}

\thanks[footnoteinfo]{This work is supported by ONR Grant No. N00014-23-1-2043.}

\author[First]{Ingyu Jang} 
\author[First]{Leila J. Bridgeman} 

\address[First]{Department of Mechanical Engineering and Material Science, Duke University, Durham, NC 27708 USA (e-mail: ij40@duke.edu; ljb48@duke.edu).}

\begin{abstract}                
In this paper, the $\cL_2$ stability of switched networks is studied based on the $QSR$-dissipativity of each agent. 
While the integration of dissipativity with switched systems has received considerable attention, most previous studies have focused on passivity, internal stability, or feedback networks involving only two agents.
This work makes two contributions:
first, the relationship between switched $QSR$-dissipativity and $\cL_2$ stability is established based on the properties of dissipativity parameters of switched systems; and second, conditions for $\cL_2$ stability of networks consisting of $QSR$-dissipative agents with switching interconnection topologies are derived. Crucially, this shows that a common storage function will exist across all modes, avoiding the need to find one, which becomes computationally taxing for large networks with many possible configurations. Numerical examples demonstrate how this can facilitate stability analysis for networked systems under arbitrary switching of swarm drones.

\end{abstract}


\end{frontmatter}

\section{Introduction}
Control over wireless networks is increasingly common, but communication lines can drop and reconnect unexpectedly, creating discrete switches in the network dynamics, so it is crucial to design controllers that are robust to these changes. 
However, a switched system may become unstable even if each mode is individually stable \citep{liberzon2003switching}. When studying internal stability, to preclude this, a shared Lyapunov function can be sought for all modes. If individual Lyapunov functions are sought for each mode, compatibility conditions must be imposed between them during mode transitions. Either way, you must solve a Lyapunov search problem for all modes simultaneously. This can become wildly computationally taxing when changes in a network's topology induces mode transitions: finding a Lyapunov function for a single large network can be taxing (especially with nonlinear agents), and this must be done simultaneously for each possible interconnection topology, of which there may be many in a large network. This paper explores how \ac{io} stability analysis and a dissipativity perspective can be used to confront this untenable computational burden.


Switched systems consist of a set of submodes, and their dynamical model is selected from this set according to a switching signal that depends on either the state or time \citep{liberzon2003switching}. Common or multiple Lyapunov function techniques have been widely used to address stability analysis and the synthesis of stabilizing controllers and switching signals \citep{branicky2002multiple,bacciotti2005invariance,mancilla2006extension,zhao2008stability,zhang2015stability}.
Other studies have introduced dwell time constraints to exploit the fact a switched system can be stable if switching is sufficiently slow \citep{hespanha1999stability,zhai2001stability}.

Switched dissipativity has attracted considerable attention in recent years due to its merits on design compositionality \citep{zakeri2022passivity}. Passivity\citep{hatanaka2015passivity} and dissipativity\citep{willems1972dissipative,hill1977stability,lozano2013dissipative} have been investigated to account for energy information in switched systems. Passivity has been analyzed using storage functions\citep{chen2005passivity,zhao2006notion,zhao2008passivity} and \ac{lmi}\citep{geromel2012passivity}. \citep{bemporad2008passivity} studied the passivity properties of discrete-time hybrid systems in the piecewise polynomial form. More general notions of incremental passivity, dissipativity, and incremental $(Q,S,R)$-dissipativity have been introduced for switched systems \citep{zhao2008dissipativity,zhao2009vector,dong2012incremental,pang2016incremental}. Dissipativity of switched systems has been established based on dwell times \citep{xiang2015dissipativity}, scattering transformations combined with a common storage function \citep{polushin2021stabilization}, \ac{lmi} conditions \citep{jungers2019dissipativeness}, and 
a decomposable supply rate is separated into a main supply rate and subsidiary supply rates\citep{liu2011decomposable}.

Despite all of this research, few works have exploited a key feature of dissipativity in a switching context: interconnections of dissipative systems are themselves dissipative and \ac{io} stable if they satisfy the 
\ac{vndt} in \citep{moylan1978stability,vidyasagar1981input}. This allows the use of only open-loop analysis of individual agents to assess the stability of the entire network. 
Most results exploiting this rely on the special case of passivity. 
Some works have employed dissipativity, but their stability theorems are primarily based on internal stability, despite the critical role of \ac{io} stability in ensuring robustness \citep{hill2022dissipativity}.
Moreover, these studies are largely confined to generic switched systems, rather than considering the underlying mechanisms that induce switching. 
In contrast, this work focuses on the relationship between $QSR$-dissipativity of agents and $\cL_2$ stability of their interconnection under switching network topologies, with the goal of creating tractible computations, even in large, changing networks.

The contributions of this paper are twofold. First, we establish a relationship between switched $QSR$-dissipativity and \ac{io} stability using the dissipativity parameters, where prior work has employed more restricted switched dissipativity definitions and required intermediary notions of gain. 
Second, we specialize the result to establish \ac{io} stability conditions to switched networked systems, where switching arises solely from changing topologies, not switching agent dynamics. These conditions serve as an extension of \ac{vndt} in \citep{moylan1978stability,vidyasagar1981input} to switched networked systems, enabling network stability analysis using open-loop characterizations of individual agents. Most critically, by using the dissipativity properties of each agent, the resulting switched networked system can be guaranteed to be \ac{io} stable under arbitrary switching signals without searching for a common storage function. Numerical examples illustrate the huge computational benefit of this approach relative to straightforwardly treating the network as a switched system, and even show a moderate benefit relative to prior work that constructs scattering functions to establish stability under arbitrary switching. Since many practical networked systems involve the addition or removal of several links, this result provides a useful condition for constructing \ac{io} stable systems under any arbitrary switching laws.

\section{Preliminaries} \label{Chap:Preliminaries}

\subsection{Notation}
The sets of real, nonnegative real, nonnegative integers, natural numbers, and natural numbers up to $n$ are denoted $\bbR$, $\bbR_+$, $\bbZ_+$, $\bbN$, and $\bbN_n$, respectively. The set of real $n\times m$ matrices is $\bbR^{n\times m}$. The set of $n\times n$ symmetric matrices is denoted by $\bbS^n$ and the negative semi-definite subset of this is $\bbS^n_-$. Asterisks $(*)$ indicate transposed blocks in symmetric matrices. The notation $\bA\prec0$ indicates that $\bA$ is negative-definite. 
The number of elements in a set $\bbA$ is denoted by $|\bbA|$. 
The $(i,j)^\thh$ block of a matrix $\bA$ is $(\bA)_{i,j}$. If $(\bA)_{i,j}{\in}\bbR^{n_i{\times} m_j}$ and $\bA{\in}\bbR^{\sum_{i=1}^N n_i {\times} \sum_{j=1}^M m_j}$, then $\bA$ is said to be in $\bbR^{N{\times} M}$ block-wise.

The set of absolute integrable functions, absolute integrable positive functions, and square integrable functions are $\cL_1$, $\cL_1^+$, and $\cL_{2}$, respectively. The vector norm, Frobenius norm, and $\cL_2$ norm are denoted by $\norm{\cdot}$, $\|\cdot\|_F$ and $\|\cdot\|_2$, respectively.  The truncation of a function $\textbf{y}(t)$ at $T$ is denoted by $\by_T(t)$, where $\by_T(t)=\by(t)$ if $t\leq T$, and $\by_T(t)=0$ otherwise. If $\|\by_T\|_2^2{=}\langle\by_T,\by_T\rangle{=}\int_0^{\infty}\by_T^T(t)\by_T(t)dt{<}\infty$ for all $T{\geq}0$, then $\by{\in}\cL_{2e}$, where $\cL_{2e}$ is the extended $\cL_2$ space.

\subsection{Dissipativity}
Consider a continuous nonlinear system $\mathscr{G}:\cL_{2e}^m\to\cL_{2e}^l$, represented by the state-space realization 
\begin{align} \label{eq:continuous_dynamics}
    \mathscr{G}:\;
        \dbx(t)=f(\bx(t),\bu(t)), \quad
        \by(t)=h(\bx(t),\bu(t)),
\end{align}
where $\bx(t)\in\bbR^n$, $\bu(t)\in\bbR^m$, and $\by(t)\in\bbR^l$ denote the states, inputs, outputs of the system, respectively.
When contextually clear, the explicit time dependency will be omitted for notational simplicity.

Dissipativity, defined below, quantifies a relationship between system inputs and outputs.
\begin{definition} [Dissipativity, \citep{lozano2013dissipative}] \label{def:Dissipativity}
    The system in \autoref{eq:continuous_dynamics} is \textit{dissipative} if there exists a storage function, $V(\bx)\geq0$, satisfying the dissipation inequality,
    \begin{align} \label{eq:dissipative}
        V(\bx(t))-V(\bx(t_0))\leq\int_{t_0}^t w(\bu,\by)dt,
    \end{align}
    for any initial condition $\bx(t_0)=\bx_0$ and any $t\geq t_0$. 
    When
    \begin{align} \label{eq:QSR}
        w(\bu,\by)=\quadsupply{\bQ}{\bS}{\bR},
    \end{align}
    the system is called \textit{QSR-dissipative}.
\end{definition}

In this paper, $\cL_2$-stability is used to define the \ac{io} stability.
\begin{definition}[$\cL_2$-stability, \citep{vidyasagar1981input}] \label{def:L2 Stable}
    The system, $\mathscr{G}:\cL_{2e}^m\to\cL_{2e}^l$, defined by \autoref{eq:continuous_dynamics} is $\cL_2$-stable if there exists $\gamma>0$ and a function $\beta(\bx)$ such that
    \begin{align} \label{eq:L2_stable}
        \|\by_T\|_2\leq\gamma\|\bu_T\|_2+\beta(\bx_0).
    \end{align}
     for all $\bu\in\cL_2^m$, $\bx_0$ and $T>0$.
     When $\mathscr{G}$ is $\cL_2$ stable, infimum of $\gamma$ is called its the \textit{$\cL_2$-gain}.
\end{definition}



\subsection{Switched Systems} \label{subChap:Siwtched_System}
A nonlinear switched system is defined by the family of $M$ different modes with the form of \autoref{eq:continuous_dynamics} as
\begin{align} \label{eq:switched_system}
    \begin{split}
        \dbx=f_{\sigma(t)}(\bx,\bu), \quad
        \by=h_{\sigma(t)}(\bx,\bu),
    \end{split}
\end{align}
where $\sigma{:}\bbR_+{\to}\bbN_M$ is the switching signal which is a piecewise constant.
This defines a switching sequence,
\begin{align} \label{eq:switching_sequence}
    \hspace{-10pt}\Sigma{=}\bigl\{\hspace{-2pt}(i_k{,}t_k){|}i_k{=}\sigma(t_k){,}
    t_{k+1}{=}\min\{t{>}t_k{|}\sigma(t){\neq}\sigma(t_k)\hspace{-1pt}\}\hspace{-2pt}\bigr\}\hspace{-1pt}_{k=0}^\infty.
    \hspace{-10pt}
\end{align}

In this paper, we assume following to ensure that the state trajectory is continuous and prevent Zeno behavior \citep{liberzon2003switching}.
\begin{assumption} \label{ass:continuous_zeno}
    The state of \autoref{eq:switched_system} does not jump at the switching instants and $\sigma(t)$ can only switch a finite number of times on any finite time interval.
\end{assumption}

The stability notion in \autoref{def:L2 Stable} applies to switched systems for a given switching signal $\sigma(t)$. If the system remains $\cL_2$ stable under any possible switching signal, it is referred to as \emph{$\cL_2$ stable under arbitrary switching}.

\subsection{Switched Networked Systems} \label{subChap:Switched_Networked_Systems}
Controlling networked systems whose communication links may arbitrarily drop or reconnect is a major challenge. 
Due to the discrete changes in communication topology, these networks exhibit switched system dynamics. 
This section reviews the dynamics of switched systems and relates switched dynamics to networks with changing topologies.

Consider $N$ agents $\scrG_p{:}\cL_{2e}^{m_p}{\to}\cL_{2e}^{l_p}$ with interconnected dynamics,
\begin{align}
    \mathscr{G}_p:\;
        \dbx_p=&f^p(\bx_p,\bu_p), \quad 
        &\by_p=h^p(\bx_p,\bu_p),\nonumber\\
    \bu_p=&\be_p+\sum_{q\in\bbN_N}\bH_{pq}^{\sigma(t)}\by_q,\quad
    &\bu=\be+\bH_{\sigma(t)}\by,\label{eq:interconnection}
\end{align}
where $\bu{=}\col(\hspace{-.5pt}\bu_p)_{p{\in}\bbN_N}\hspace{-1pt}$, $\by{=}\col(\hspace{-.5pt}\by_p)_{p{\in}\bbN_N}\hspace{-1pt}$, and $\be{=}\col(\hspace{-.5pt}\be_p)_{p{\in}\bbN_N}\hspace{-1pt}$. The matrix $\bH_{\sigma(t)}$ represents the communication topology among agents, whose $(p,q)^\thh$ block, $\bH_{pq}^{\sigma(t)}{\in}\bbR^{m_p\times l_q}$, is governed by switching signal, $\sigma(t){\in}\bbN_{M}$. 
This switching signal defines the same switching sequence as in \autoref{eq:switching_sequence}.
Therefore, the networked system $\hatscrG{:}\be{\to}\by$, described by \autoref{eq:interconnection}, is a special case of \autoref{eq:switched_system}, which we call a \textit{switched networked system}.


\section{Switched Dissipativity and $\cL_2$ Stability Analysis}
In this section, the notion of dissipativity and $\cL_2$-stability for non-switched systems is extended to switched systems and switched networked systems.
\citep{zhao2008dissipativity} first introduced the switched dissipativity with cross supply rate.
However, that definition includes properties directly associated with the internal stability of the switched system, which may restrict the general applicability of the switched dissipativity framework.
In this paper, a relaxed version of the definition in \citep{zhao2008dissipativity} is adopted to avoid this.
In addition, conditions ensuring the $\cL_2$ stability under arbitrary switching of the switched and switched networked system are established.

\subsection{Switched Dissipativity} \label{subChap:Switched_dissipativity}
The concept of switched dissipativity was first defined in \citep[Def. 3.3]{zhao2008dissipativity}, where it was defined using positive-definite multiple storage functions and 3 conditions.
The first condition specifies the dissipativity for the active mode, the second condition describes the dissipativity for the inactive modes, and the last condition characterizes the properties of the active and cross supply rates for the specific inputs.
Since the positive-definiteness of storage functions and the last condition were primarily used to prove the internal stability of switched systems, this formulation can sometimes limit the applicability of the switched dissipativity definition to other stability notions.
Therefore, in \autoref{def:switched-dissipativity} we adopt a relaxed version of \citep[Def. 3.3]{zhao2008dissipativity} to define the dissipativity of switched systems.

\begin{definition} [Switched Dissipativity] \label{def:switched-dissipativity}
    The system in \autoref{eq:switched_system} is said to be \textit{switched dissipative} under the switching sequence $\Sigma$ if there exist continuous functions $V_i(\bx)\geq0$, so-called storage functions, active supply rates $w_i^i(\bu,\by)$, and cross supply rates $w_j^i(\bx,\bu,\by,t)$ for $i,j\in\bbN_M$ and $j\neq i$ such that
    \begin{align}
        V_{i_k}(\bx(t))-V_{i_k}(\bx(s))&\leq\int_s^tw_{i_k}^{i_k}(\bu,\by)d\tau  \numberthis\label{eq:active_dissipativity}\\
        V_{j}(\bx(t))-V_{j}(\bx(s))&\leq\int_s^tw_{j}^{i_k}(\bx,\bu,\by,\tau)d\tau \numberthis\label{eq:inactive_dissipativity}
    \end{align}
    for all $k\in\bbN$, $t_k\leq s\leq t< t_{k+1}$, and $(i_k,t_k)\in\Sigma$. 
    When
    \begin{align} \label{eq:quadratic_supply_rate}
        w_{i}^{i}(\bu,\by)=\quadsupply{\bQ_i}{\bS_i}{\bR_i},\quad\forall i\in\bbN_M
    \end{align}
    the system is said to be \textit{switched QSR-dissipative}.
\end{definition}

\subsection{\texorpdfstring{$\cL_2$}{L2} Stability with Common Storage Function} \label{subChap:L2_stability}
The $\cL_2$ stability of switched systems has been studied in the literature.
For instance, \citep{zhao2008stability,zhang2015stability} established $\cL_2$ stability using multiple Lyapunov functions, while \citep{zhao2009vector} investigated the relationship between the individual $\cL_2$ gain of switched systems' modes, so-called vector $\cL_2$ gain, and their internal stability.
In contrast, our stability theorem, \autoref{thm:global_switched_stability}, discusses the conditions under which $\cL_2$ stability can be inferred from the properties of the swithced $QSR$-dissipativity parameters and storage functions. Moreover, our theorem is built from the less restrictive dissipativity notion from \autoref{def:switched-dissipativity}.

\begin{theorem} \label{thm:global_switched_stability}
    Consider a switched system in \autoref{eq:switched_system}. If, for any $i\in\bbN_M$, the switched system is switched $QSR$-dissipative with a common storage function $V_i(\bx)=V(\bx)$ and satisfies $\bQ_i\prec0$, then the system has a common supply rate and is $\cL_2$ stable under arbitrary switching.
\end{theorem}
\begin{pf}
    With a common storage function $V(\bx)$, \autoref{eq:active_dissipativity} gives
    \begin{align*}
        V(\bx(t))-V(\bx(s))&\leq\int_s^t\quadsupply{\bQ_{i_k}}{\bS_{i_k}}{\bR_{i_k}}d\tau
    \end{align*}
    for all $k\in\bbN$, $t_k\leq s\leq t< t_{k+1}$, and $(i_k,t_k)\in\Sigma$.
    Since $\bQ_i\prec0$, for each $i\in\bbN_M$, there exists $\epsilon_i>0$ such that $\bQ_i+\epsilon_i\bI\prec0$.
    By Young's relation, the quadratic supply rate of $i^\thh$ mode satisfies that
    \begin{align*}
        &\by^T\bQ_i\by{+}\by^T\bS_i\bu{+}\bu^T\bS_i^T\by{+}\bu^T\bR_i\bu \\
        &\quad{\leq}\by^T(\bQ_i{+}\epsilon_i\bI)\by{+}\bu^T\Big(\frac{1}{\epsilon_i}\bS_i^T\bS_i{+}\bR_i\Big)\bu
    \end{align*}
    Define $q=-\max_{i\in\bbN_M}\lambda_{\max}(\bQ_i+\epsilon_i\bI)$, which is always positive, and $r=\max_{i\in\bbN_M}\lambda_{\max}(\frac{1}{\epsilon_i}\bS_i^T\bS_i+\bR_i)$.
    Then
    \begin{align*}
        V(\bx(t))-V(\bx(s))&\leq-q\int_s^t\by^T\by d\tau+r\int_s^t\bu^T\bu d\tau,
    \end{align*}
    for all $k\in\bbN$, $t_k\leq s\leq t< t_{k+1}$, and $(i_k,t_k)\in\Sigma$, which shows that the switched system is switched $QSR$-dissipative with a common supply rate characterized by $\bQ=-q\bI$, $\bS=\bzero$, and $\bR=r\bI$.
    In addition, since $V(\bx)\geq0$ for all $\bx\in\bbR^n$ and $V(\bx)$ is continuous, 
    \begin{align*}
        -V(\bx(0))&\leq-q\int_0^T\by^T\by d\tau+r\int_0^T\bu^T\bu d\tau,\quad\forall T\geq0.
    \end{align*}
    Therefore, $\cL_2$ stable under arbitrary switching.
    $\hfill\blacksquare$
\end{pf}
\subsection{Stability and Storage Functions}


A common storage function satisfying \autoref{thm:global_switched_stability} does not always exist, motivating various techniques to establish stability conditions using multiple Lyapunov or storage functions. The switched networked systems follow the same dynamics as switched systems, but constructing a Lyapunov function for a single large-scale network, particularly with nonlinear agents, is taxing. Our main result, \autoref{thm:global_switched_stability}, shows how to avoid this difficulty in switched networked systems with dissipative agents.


\section{$\cL_2$ Stability of Switched Networked Systems}\label{subChap:L2_stable_switched_networked_systems}

This section establishes the stability of a switched networked system based on the local dissipativity properties of individual agents.
The open-loop properties of each dissipative agent play a critical role in determining overall stability under arbitrary switching, allowing us to construct a common storage function from the individual storage functions of each agent. Hence, the difficulty of constructing such common storage function can be avoided for switched networked systems. The result is sufficient conditions for stability, similar to the \ac{vndt} for non-switched networked systems.
%
%
\begin{theorem} \label{thm:stability_network_nonswitched_agents}
    Consider the networked system $\hatscrG:\be\to\bu$ of $N$ agents defined in \autoref{eq:interconnection}. 
    Assume that each agent is dissipative with storage function $V_p(\bx_p)$, and dissipativity parameters $\bQ_p$, $\bS_p$, and $\bR_p$.
    Then the switched networked system is $\cL_2$ stable under arbitrary switching if $\hatbQ_{i}{\prec}0$ for all $i{\in}\bbN_{M}$, where
    \begin{align*}
        \hatbQ_{i}&{=}\bQ{+}\bS\bH_{i}{+}\bH_{i}^T\bS{+}\bH_{i}^T\bR\bH_{i}, \\
        \bQ&{=}\diag(\bQ_p)_{p\in\bbN_N},\,\bS{=}\diag(\bS_p)_{p\in\bbN_N},\,\bR{=}\diag(\bR_p)_{p\in\bbN_N}.
    \end{align*}
\end{theorem}

\begin{pf}
    From the $QSR$-dissipativity property of each agent, it follows that
    \begin{align*}
        \sum_{p\in\bbN_N}\bracketl{V_p(\bx_p(T))-V_p(\bx_p(0))}\leq\int_0^T\quadsupply{\bQ}{\bS}{\bR}dt.
    \end{align*}
    Using \autoref{eq:interconnection}, we obtain
    \begin{align*}
        \hspace{-2pt}\int_0^T\hspace{-4pt}\quadsupply{\bQ}{\bS}{\bR}\hspace{-3pt}dt 
        &{=}\hspace{-5pt}\int_0^T\hspace{-4pt}\quadsupplyinput{\bQ}{\bS}{\bR}{\by}{\be{+}\bH_{\sigma(t)}\hspace{-1.5pt}\by}\hspace{-3pt}dt \\
        &{=}\hspace{-5pt}\int_0^T\hspace{-4pt}\quadsupplyinput{\hatbQ_{\sigma(t)}}{\hatbS_{\sigma(t)}}{\hatbR_{\sigma(t)}}{\by}{\be}\hspace{-3pt}dt,
    \end{align*}
    where 
    $\hatbS_{i}{=}\bS{+}\bH_{\hati}^T\bR$, and $\hatbR_{\hati}{=}\bR$. It follows that the system is switched $QSR$-dissipative with a common storage function $\sum_{p\in\bbN_N}V_p(\bx_p)$. Therefore, from \autoref{thm:global_switched_stability}, the system is $\cL_2$ stable under arbitrary switching.
    $\hfill\blacksquare$
\end{pf}

\subsection{Utility of \autoref{thm:stability_network_nonswitched_agents}} \label{subChap:utility_of_thm2}

\autoref{thm:stability_network_nonswitched_agents} implies that the $\cL_2$ stability under arbitrary switching of a switched networked system can be determined directly from the local dissipativity parameters of agents and switching topology modes; there is no need to seek out the common storage function, which would otherwise be computationally heavy, involving all topologies and agents, while \autoref{thm:stability_network_nonswitched_agents} separates agent-level and interconnection constraints for stability.
A common approach to verify the $\cL_2$ stability under arbitrary switching of such systems is to construct a switched networked system and check whether it admits a common Lyapunov function, or whether it is dissipative with a common storage function and a common supply rate.
Another approach is to employ a scattering function that modifies the input and output of the switched networked system to obtain a common storage function and a common prescribed supply rate \cite[Thm. 1]{polushin2021stabilization}.
In contrast, \autoref{thm:stability_network_nonswitched_agents} is advantageous because it establishes the $\cL_2$ stability under arbitrary switching solely from the dissipativity information of individual agents, without requiring the construction of a global switched networked system or additional transformation.

\section{NUMERICAL EXAMPLE} \label{Chap:Numerical Example}

\begin{figure}
    \centering
    \includegraphics[width = 0.45\textwidth]{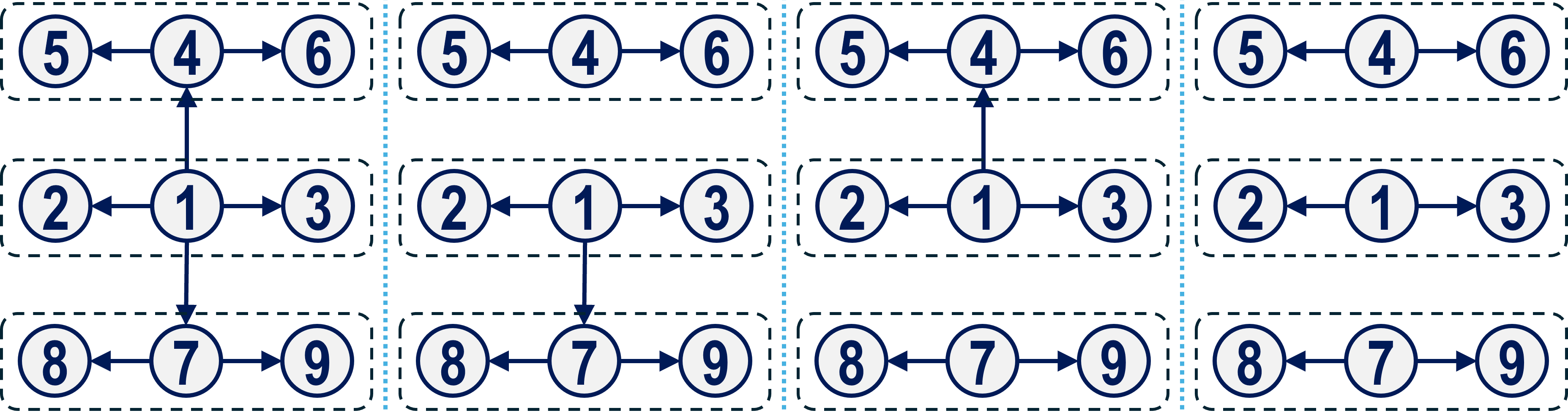}
    \caption{Switching modes of \ac{uav}s' controllers}
    \label{fig:network_interconnection}
\end{figure}
We investigate the stability of a network comprising 9 linearized \acp{uav} within the switched-dissipativity framework.
The network consists of 18 agents, including feedback controllers embedded within each \ac{uav}.
The objective in this example is to determine whether the switched network is $\cL_2$ stable under arbitrary switching signals, assuming that the external inputs belong to the $\cL_2$ function space.
The analysis considers a switched networked system subject to arbitrary switching events.


\subsection{Swicthed Network of Unmanned Aerial Vehicles} \label{subsubChap:Network_with_a_switched_topology}

We model the \acp{uav} following the linearized \ac{uav} model in \citep{luis2016design}, using their dynamics matrices $\bA_p\in\bbR^{12\times12}$ and $\bB_p\in\bbR^{12\times4}$ for all $p\in\bbN_9$, and physical parameters, except for the mass and wing length.
These are independently varied according to a uniform random distribution over $[\frac{2}{3},\,\frac{4}{3}]$ of their values reported in \citep{luis2016design}.
The input and output of each \ac{uav} are defined as the 4 rotors' rotational speeds, $\bu_p\in\bbR^{4}$, and its state vector, $\bx_p\in\bbR^{12}$, respectively.
For all $p\in\bbN_9$, the \ac{lqr} controllers $\bK_p\in\bbR^{4\times12}$ are designed with weighting matrices $\bQ_{lqr}=\diag(100\bI_6,10\bI_6)$ and $\bR_{lqr}=\bI_6$.

Each \ac{uav} communicates its state information with other \acp{uav} exclusively through the controllers. 
The basic interconnection structures $\hatbH_{i}$ of \acp{uav} with feedback controllers is expressed as
\begin{align*}
    \hatbH_{i_0}=\begin{bmatrix}
        \bzero & -\bI \\ \bH_{i}^c & \bzero
    \end{bmatrix}\in\bbR^{144\times144},\quad\forall i\in\bbN_{M},
\end{align*}
where $M=4$.
When the \acp{uav} are partitioned into 3 subgroups, $(1,2,3)$, $(4,5,6)$, and $(7,8,9)$, with switching interconnection modes given by
\setlength{\arraycolsep}{0.5pt}
\begin{align*}
    \bH_1^c{=}\hspace{-3pt}\bracketl{\hspace{-2pt}\begin{array}{ccc}
        \bH & \bzero & \bzero \\
        \widetilde{\bH} & \bH & \bzero \\
        \widetilde{\bH} & \bzero & \bH
    \end{array}\hspace{-2pt}}\hspace{-4pt},
    \bH_2^c{=}\hspace{-3pt}\bracketl{\hspace{-2pt}\begin{array}{ccc}
        \bH & \bzero & \bzero \\
        \bzero & \bH & \bzero \\
        \widetilde{\bH} & \bzero & \bH
    \end{array}\hspace{-2pt}}\hspace{-4pt},
    \bH_3^c{=}\hspace{-3pt}\bracketl{\hspace{-2pt}\begin{array}{ccc}
        \bH & \bzero & \bzero \\
        \widetilde{\bH} & \bH & \bzero \\
        \bzero & \bzero & \bH
    \end{array}\hspace{-2pt}}\hspace{-4pt},
    \bH_4^c{=}\hspace{-3pt}\bracketl{\hspace{-1pt}\begin{array}{ccc}
        \bH & \bzero & \bzero \\
        \bzero & \bH & \bzero \\
        \bzero & \bzero & \bH
    \end{array}\hspace{-2pt}}
\end{align*}
where 
\setlength{\arraycolsep}{2pt}
\begin{align*}
    \bH=\begin{bmatrix}
        \bI & \bzero & \bzero \\
        -\bI & \bI & \bzero \\
        -\bI & \bzero & \bI
    \end{bmatrix},\text{ and }
    \widetilde{\bH}=\begin{bmatrix}
        -\bI & \bzero & \bzero \\
        \bzero & \bzero & \bzero \\
        \bzero & \bzero & \bzero
    \end{bmatrix}.
\end{align*}
The corresponding switched network topologies are illustrated in \autoref{fig:network_interconnection}.

Each agent's combined plant and the controller dynamics are
\begin{align} \label{eq:uav_dynamics}
    \dbx_p{=}\bA_p\bx_p{+}\bB_p\bu_p,
    \bu_p{=}{-}\bK_p\bigg(\bx_p{+}\hspace{-3pt}\sum_{q{\in}\bbN_9}\hspace{-2pt}(\bH_{\sigma(t)}^c)_{p,q}\bigg).
\end{align}
Following \autoref{thm:stability_network_nonswitched_agents}, $\cL_2$ stability of the changing network is verified by solving
\begin{align*}
    \begin{bmatrix}
        \bA_p^T\bP_p{+}\bP_p\bA_p{-}\bQ_p & \bP_p\bB_p{-}\bS_p \\
        * & {-}\bR_p
    \end{bmatrix}{\preceq}0 \\
    {-}\tilbR_p{-}\tilbS_p^T\bK_p{-}\bK_p^T\tilbS_p{-}\bK_p^T\tilbQ_p\bK_p{\preceq}0 \\
    \bQ+\bS\bH_{i}+\bH_{i}^T\bS^T+\bH_{i}^T\bR\bH_{i}{\prec}0
\end{align*}
for all $p\in\bbN_9$ and $i\in\bbN_4$, where $\bQ=\diag(\diag(\bQ_p)_{p\in\bbN_9},\\\diag(\tilbQ_p)_{p\in\bbN_9})$, $\bS{=}\diag(\diag(\bS_p)_{p\in\bbN_9},\diag(\tilbS_p)_{p\in\bbN_9})$, $\bR=\diag(\diag(\bR_p)_{p\in\bbN_9},\diag(\tilbR_p)_{p\in\bbN_9})$
and $(\bQ_p,\bS_p,\bR_p)$ and $(\tilbQ_p,\tilbS_p,\tilbR_p)$ denote the dissipativity parameters for the $p^\thh$ \ac{uav} and its controller, respectively.

\subsection{Results}
The feasibility problems described in \autoref{subsubChap:Network_with_a_switched_topology} are solved using MOSEK \citep{aps2019mosek}, YALMIP \citep{lofberg2004yalmip}, and MATLAB.
The problem is feasible, which implies that the corresponding switched networked system is $\cL_2$ stable under arbitrary switching. 
In the following, we further verify this result by examining the system response to disturbances belonging to the $\cL_2$ function space.

Every component of the state and input rpm of each agent is subjected to a disturbance belonging to the $\cL_2$ function space. 
For each component, the disturbances are randomly selected from 3 functions, $f_1(t)=c_1t^2e^{-t}$, $f_2(t)=c_2\frac{\sin(t)}{t}$, and $f_3(t)=c_3\frac{1}{1+t}$, where each function is scaled by a constant factor, $c_i$ for $i\in\bbN_3$, such that its $\cL_2$ norm is approximately 1.
Furthermore, the disturbances applied to the input rpm are scaled by a factor of 1000 relative to those applied to the states, since the magnitudes of rpm values are significantly larger than those of the states.


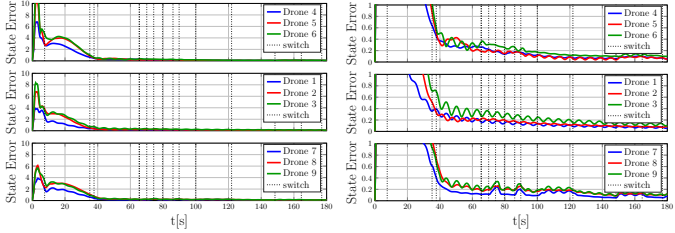
\begin{figure}
\centering
    \begin{subfigure}[t]{0.24\textwidth}
        \centering
        \resizebox{\textwidth}{!}{
%
%
\begin{tikzpicture}

\begin{axis}[%
width=5.016in,
height=0.986in,
at={(0.54in,1.714in)},
scale only axis,
xmin=0,
xmax=180,
xtick={0,20,40,60,80,100,120,140,160,180},
xticklabels={\empty},
ymin=0,
ymax=10,
ylabel style={font=\color{white!15!black}},
ylabel={State Error},
axis background/.style={fill=white},
xmajorgrids,
ymajorgrids,
legend style={legend cell align=left, align=left, draw=white!15!black},
title style={font=\Huge},xlabel style={font={\LARGE}},ylabel style={font=\LARGE},legend style={font=\large},
]
\addplot [color=blue, line width=2pt]
  table[x=t, y=e, col sep=comma]
  {Figures_tex/Data/Arbitrary_Drone_1.csv};
\addlegendentry{Drone 1}

\addplot [color=red, line width=2pt]
  table[x=t, y=e, col sep=comma]
  {Figures_tex/Data/Arbitrary_Drone_2.csv};
\addlegendentry{Drone 2}

\addplot [color=green!60!black, line width=2pt]
  table[x=t, y=e, col sep=comma]
  {Figures_tex/Data/Arbitrary_Drone_3.csv};
\addlegendentry{Drone 3}

\addplot [color=black, dotted, line width=1.0pt]
  table[row sep=crcr]{%
7.54166666666667	0\\
7.54166666666667	100\\
};
\addlegendentry{switch}

\addplot [color=black, dotted, line width=1.0pt, forget plot]
  table[row sep=crcr]{%
35.1666666666667	0\\
35.1666666666667	100\\
};
\addplot [color=black, dotted, line width=1.0pt, forget plot]
  table[row sep=crcr]{%
37.75	0\\
37.75	100\\
};
\addplot [color=black, dotted, line width=1.0pt, forget plot]
  table[row sep=crcr]{%
65.4166666666667	0\\
65.4166666666667	100\\
};
\addplot [color=black, dotted, line width=1.0pt, forget plot]
  table[row sep=crcr]{%
65.4583333333333	0\\
65.4583333333333	100\\
};
\addplot [color=black, dotted, line width=1.0pt, forget plot]
  table[row sep=crcr]{%
70	0\\
70	100\\
};
\addplot [color=black, dotted, line width=1.0pt, forget plot]
  table[row sep=crcr]{%
74.25	0\\
74.25	100\\
};
\addplot [color=black, dotted, line width=1.0pt, forget plot]
  table[row sep=crcr]{%
79.7083333333333	0\\
79.7083333333333	100\\
};
\addplot [color=black, dotted, line width=1.0pt, forget plot]
  table[row sep=crcr]{%
86.0416666666667	0\\
86.0416666666667	100\\
};
\addplot [color=black, dotted, line width=1.0pt, forget plot]
  table[row sep=crcr]{%
89.7916666666667	0\\
89.7916666666667	100\\
};
\addplot [color=black, dotted, line width=1.0pt, forget plot]
  table[row sep=crcr]{%
99.9583333333333	0\\
99.9583333333333	100\\
};
\addplot [color=black, dotted, line width=1.0pt, forget plot]
  table[row sep=crcr]{%
121.875	0\\
121.875	100\\
};
\addplot [color=black, dotted, line width=1.0pt, forget plot]
  table[row sep=crcr]{%
148.416666666667	0\\
148.416666666667	100\\
};
\addplot [color=black, dotted, line width=1.0pt, forget plot]
  table[row sep=crcr]{%
164.541666666667	0\\
164.541666666667	100\\
};
\addplot [color=black, dotted, line width=1.0pt, forget plot]
  table[row sep=crcr]{%
177	0\\
177	100\\
};
\end{axis}

\begin{axis}[%
width=5.016in,
height=0.986in,
at={(0.54in,2.914in)},
scale only axis,
xmin=0,
xmax=180,
xtick={0,20,40,60,80,100,120,140,160,180},
xticklabels={\empty},
ymin=0,
ymax=10,
ylabel style={font=\color{white!15!black}},
ylabel={State Error},
axis background/.style={fill=white},
xmajorgrids,
ymajorgrids,
legend style={legend cell align=left, align=left, draw=white!15!black},
title style={font=\Huge},xlabel style={font={\LARGE}},ylabel style={font=\LARGE},legend style={font=\large},
]
\addplot [color=blue, line width=2pt]
  table[x=t, y=e, col sep=comma]
  {Figures_tex/Data/Arbitrary_Drone_4.csv};
\addlegendentry{Drone 4}

\addplot [color=red, line width=2pt]
  table[x=t, y=e, col sep=comma]
  {Figures_tex/Data/Arbitrary_Drone_5.csv};
\addlegendentry{Drone 5}

\addplot [color=green!60!black, line width=2pt]
  table[x=t, y=e, col sep=comma]
  {Figures_tex/Data/Arbitrary_Drone_6.csv};
\addlegendentry{Drone 6}

\addplot [color=black, dotted, line width=1.0pt]
  table[row sep=crcr]{%
7.54166666666667	0\\
7.54166666666667	100\\
};
\addlegendentry{switch}

\addplot [color=black, dotted, line width=1.0pt, forget plot]
  table[row sep=crcr]{%
35.1666666666667	0\\
35.1666666666667	100\\
};
\addplot [color=black, dotted, line width=1.0pt, forget plot]
  table[row sep=crcr]{%
37.75	0\\
37.75	100\\
};
\addplot [color=black, dotted, line width=1.0pt, forget plot]
  table[row sep=crcr]{%
65.4166666666667	0\\
65.4166666666667	100\\
};
\addplot [color=black, dotted, line width=1.0pt, forget plot]
  table[row sep=crcr]{%
65.4583333333333	0\\
65.4583333333333	100\\
};
\addplot [color=black, dotted, line width=1.0pt, forget plot]
  table[row sep=crcr]{%
70	0\\
70	100\\
};
\addplot [color=black, dotted, line width=1.0pt, forget plot]
  table[row sep=crcr]{%
74.25	0\\
74.25	100\\
};
\addplot [color=black, dotted, line width=1.0pt, forget plot]
  table[row sep=crcr]{%
79.7083333333333	0\\
79.7083333333333	100\\
};
\addplot [color=black, dotted, line width=1.0pt, forget plot]
  table[row sep=crcr]{%
86.0416666666667	0\\
86.0416666666667	100\\
};
\addplot [color=black, dotted, line width=1.0pt, forget plot]
  table[row sep=crcr]{%
89.7916666666667	0\\
89.7916666666667	100\\
};
\addplot [color=black, dotted, line width=1.0pt, forget plot]
  table[row sep=crcr]{%
99.9583333333333	0\\
99.9583333333333	100\\
};
\addplot [color=black, dotted, line width=1.0pt, forget plot]
  table[row sep=crcr]{%
121.875	0\\
121.875	100\\
};
\addplot [color=black, dotted, line width=1.0pt, forget plot]
  table[row sep=crcr]{%
148.416666666667	0\\
148.416666666667	100\\
};
\addplot [color=black, dotted, line width=1.0pt, forget plot]
  table[row sep=crcr]{%
164.541666666667	0\\
164.541666666667	100\\
};
\addplot [color=black, dotted, line width=1.0pt, forget plot]
  table[row sep=crcr]{%
177	0\\
177	100\\
};
\end{axis}

\begin{axis}[%
width=5.016in,
height=0.986in,
at={(0.54in,0.515in)},
scale only axis,
xmin=0,
xmax=180,
xlabel style={font=\color{white!15!black}},
xlabel={t[s]},
ymin=0,
ymax=10,
ylabel style={font=\color{white!15!black}},
ylabel={State Error},
axis background/.style={fill=white},
xmajorgrids,
ymajorgrids,
legend style={legend cell align=left, align=left, draw=white!15!black},
title style={font=\Huge},xlabel style={font={\LARGE}},ylabel style={font=\LARGE},legend style={font=\large},
]
\addplot [color=blue, line width=2pt]
  table[x=t, y=e, col sep=comma]
  {Figures_tex/Data/Arbitrary_Drone_7.csv};
\addlegendentry{Drone 7}

\addplot [color=red, line width=2pt]
  table[x=t, y=e, col sep=comma]
  {Figures_tex/Data/Arbitrary_Drone_8.csv};
\addlegendentry{Drone 8}

\addplot [color=green!60!black, line width=2pt]
  table[x=t, y=e, col sep=comma]
  {Figures_tex/Data/Arbitrary_Drone_9.csv};
\addlegendentry{Drone 9}

\addplot [color=black, dotted, line width=1.0pt]
  table[row sep=crcr]{%
7.54166666666667	0\\
7.54166666666667	100\\
};
\addlegendentry{switch}

\addplot [color=black, dotted, line width=1.0pt, forget plot]
  table[row sep=crcr]{%
35.1666666666667	0\\
35.1666666666667	100\\
};
\addplot [color=black, dotted, line width=1.0pt, forget plot]
  table[row sep=crcr]{%
37.75	0\\
37.75	100\\
};
\addplot [color=black, dotted, line width=1.0pt, forget plot]
  table[row sep=crcr]{%
65.4166666666667	0\\
65.4166666666667	100\\
};
\addplot [color=black, dotted, line width=1.0pt, forget plot]
  table[row sep=crcr]{%
65.4583333333333	0\\
65.4583333333333	100\\
};
\addplot [color=black, dotted, line width=1.0pt, forget plot]
  table[row sep=crcr]{%
70	0\\
70	100\\
};
\addplot [color=black, dotted, line width=1.0pt, forget plot]
  table[row sep=crcr]{%
74.25	0\\
74.25	100\\
};
\addplot [color=black, dotted, line width=1.0pt, forget plot]
  table[row sep=crcr]{%
79.7083333333333	0\\
79.7083333333333	100\\
};
\addplot [color=black, dotted, line width=1.0pt, forget plot]
  table[row sep=crcr]{%
86.0416666666667	0\\
86.0416666666667	100\\
};
\addplot [color=black, dotted, line width=1.0pt, forget plot]
  table[row sep=crcr]{%
89.7916666666667	0\\
89.7916666666667	100\\
};
\addplot [color=black, dotted, line width=1.0pt, forget plot]
  table[row sep=crcr]{%
99.9583333333333	0\\
99.9583333333333	100\\
};
\addplot [color=black, dotted, line width=1.0pt, forget plot]
  table[row sep=crcr]{%
121.875	0\\
121.875	100\\
};
\addplot [color=black, dotted, line width=1.0pt, forget plot]
  table[row sep=crcr]{%
148.416666666667	0\\
148.416666666667	100\\
};
\addplot [color=black, dotted, line width=1.0pt, forget plot]
  table[row sep=crcr]{%
164.541666666667	0\\
164.541666666667	100\\
};
\addplot [color=black, dotted, line width=1.0pt, forget plot]
  table[row sep=crcr]{%
177	0\\
177	100\\
};

\end{axis}
\end{tikzpicture}%
}
    \end{subfigure}
    \begin{subfigure}[t]{0.24\textwidth}
        \centering
        \resizebox{\textwidth}{!}{
%
%
\begin{tikzpicture}

\begin{axis}[%
width=5.016in,
height=0.986in,
at={(0.54in,1.714in)},
scale only axis,
xmin=0,
xmax=180,
xtick={0,20,40,60,80,100,120,140,160,180},
xticklabels={\empty},
ymin=0,
ymax=1,
ylabel style={font=\color{white!15!black}},
ylabel={State Error},
axis background/.style={fill=white},
xmajorgrids,
ymajorgrids,
legend style={legend cell align=left, align=left, draw=white!15!black},
title style={font=\Huge},xlabel style={font={\LARGE}},ylabel style={font=\LARGE},legend style={font=\large},
]
\addplot [color=blue, line width=2pt]
  table[x=t, y=e, col sep=comma]
  {Figures_tex/Data/Arbitrary_Drone_1.csv};
\addlegendentry{Drone 1}

\addplot [color=red, line width=2pt]
  table[x=t, y=e, col sep=comma]
  {Figures_tex/Data/Arbitrary_Drone_2.csv};
\addlegendentry{Drone 2}

\addplot [color=green!60!black, line width=2pt]
  table[x=t, y=e, col sep=comma]
  {Figures_tex/Data/Arbitrary_Drone_3.csv};
\addlegendentry{Drone 3}

\addplot [color=black, dotted, line width=1.0pt]
  table[row sep=crcr]{%
7.54166666666667	0\\
7.54166666666667	100\\
};
\addlegendentry{switch}

\addplot [color=black, dotted, line width=1.0pt, forget plot]
  table[row sep=crcr]{%
35.1666666666667	0\\
35.1666666666667	100\\
};
\addplot [color=black, dotted, line width=1.0pt, forget plot]
  table[row sep=crcr]{%
37.75	0\\
37.75	100\\
};
\addplot [color=black, dotted, line width=1.0pt, forget plot]
  table[row sep=crcr]{%
65.4166666666667	0\\
65.4166666666667	100\\
};
\addplot [color=black, dotted, line width=1.0pt, forget plot]
  table[row sep=crcr]{%
65.4583333333333	0\\
65.4583333333333	100\\
};
\addplot [color=black, dotted, line width=1.0pt, forget plot]
  table[row sep=crcr]{%
70	0\\
70	100\\
};
\addplot [color=black, dotted, line width=1.0pt, forget plot]
  table[row sep=crcr]{%
74.25	0\\
74.25	100\\
};
\addplot [color=black, dotted, line width=1.0pt, forget plot]
  table[row sep=crcr]{%
79.7083333333333	0\\
79.7083333333333	100\\
};
\addplot [color=black, dotted, line width=1.0pt, forget plot]
  table[row sep=crcr]{%
86.0416666666667	0\\
86.0416666666667	100\\
};
\addplot [color=black, dotted, line width=1.0pt, forget plot]
  table[row sep=crcr]{%
89.7916666666667	0\\
89.7916666666667	100\\
};
\addplot [color=black, dotted, line width=1.0pt, forget plot]
  table[row sep=crcr]{%
99.9583333333333	0\\
99.9583333333333	100\\
};
\addplot [color=black, dotted, line width=1.0pt, forget plot]
  table[row sep=crcr]{%
121.875	0\\
121.875	100\\
};
\addplot [color=black, dotted, line width=1.0pt, forget plot]
  table[row sep=crcr]{%
148.416666666667	0\\
148.416666666667	100\\
};
\addplot [color=black, dotted, line width=1.0pt, forget plot]
  table[row sep=crcr]{%
164.541666666667	0\\
164.541666666667	100\\
};
\addplot [color=black, dotted, line width=1.0pt, forget plot]
  table[row sep=crcr]{%
177	0\\
177	100\\
};
\end{axis}

\begin{axis}[%
width=5.016in,
height=0.986in,
at={(0.54in,2.914in)},
scale only axis,
xmin=0,
xmax=180,
xtick={0,20,40,60,80,100,120,140,160,180},
xticklabels={\empty},
ymin=0,
ymax=1,
ylabel style={font=\color{white!15!black}},
ylabel={State Error},
axis background/.style={fill=white},
xmajorgrids,
ymajorgrids,
legend style={legend cell align=left, align=left, draw=white!15!black},
title style={font=\Huge},xlabel style={font={\LARGE}},ylabel style={font=\LARGE},legend style={font=\large},
]
\addplot [color=blue, line width=2pt]
  table[x=t, y=e, col sep=comma]
  {Figures_tex/Data/Arbitrary_Drone_4.csv};
\addlegendentry{Drone 4}

\addplot [color=red, line width=2pt]
  table[x=t, y=e, col sep=comma]
  {Figures_tex/Data/Arbitrary_Drone_5.csv};
\addlegendentry{Drone 5}

\addplot [color=green!60!black, line width=2pt]
  table[x=t, y=e, col sep=comma]
  {Figures_tex/Data/Arbitrary_Drone_6.csv};
\addlegendentry{Drone 6}

\addplot [color=black, dotted, line width=1.0pt]
  table[row sep=crcr]{%
7.54166666666667	0\\
7.54166666666667	100\\
};
\addlegendentry{switch}

\addplot [color=black, dotted, line width=1.0pt, forget plot]
  table[row sep=crcr]{%
35.1666666666667	0\\
35.1666666666667	100\\
};
\addplot [color=black, dotted, line width=1.0pt, forget plot]
  table[row sep=crcr]{%
37.75	0\\
37.75	100\\
};
\addplot [color=black, dotted, line width=1.0pt, forget plot]
  table[row sep=crcr]{%
65.4166666666667	0\\
65.4166666666667	100\\
};
\addplot [color=black, dotted, line width=1.0pt, forget plot]
  table[row sep=crcr]{%
65.4583333333333	0\\
65.4583333333333	100\\
};
\addplot [color=black, dotted, line width=1.0pt, forget plot]
  table[row sep=crcr]{%
70	0\\
70	100\\
};
\addplot [color=black, dotted, line width=1.0pt, forget plot]
  table[row sep=crcr]{%
74.25	0\\
74.25	100\\
};
\addplot [color=black, dotted, line width=1.0pt, forget plot]
  table[row sep=crcr]{%
79.7083333333333	0\\
79.7083333333333	100\\
};
\addplot [color=black, dotted, line width=1.0pt, forget plot]
  table[row sep=crcr]{%
86.0416666666667	0\\
86.0416666666667	100\\
};
\addplot [color=black, dotted, line width=1.0pt, forget plot]
  table[row sep=crcr]{%
89.7916666666667	0\\
89.7916666666667	100\\
};
\addplot [color=black, dotted, line width=1.0pt, forget plot]
  table[row sep=crcr]{%
99.9583333333333	0\\
99.9583333333333	100\\
};
\addplot [color=black, dotted, line width=1.0pt, forget plot]
  table[row sep=crcr]{%
121.875	0\\
121.875	100\\
};
\addplot [color=black, dotted, line width=1.0pt, forget plot]
  table[row sep=crcr]{%
148.416666666667	0\\
148.416666666667	100\\
};
\addplot [color=black, dotted, line width=1.0pt, forget plot]
  table[row sep=crcr]{%
164.541666666667	0\\
164.541666666667	100\\
};
\addplot [color=black, dotted, line width=1.0pt, forget plot]
  table[row sep=crcr]{%
177	0\\
177	100\\
};
\end{axis}

\begin{axis}[%
width=5.016in,
height=0.986in,
at={(0.54in,0.515in)},
scale only axis,
xmin=0,
xmax=180,
xlabel style={font=\color{white!15!black}},
xlabel={t[s]},
ymin=0,
ymax=1,
ylabel style={font=\color{white!15!black}},
ylabel={State Error},
axis background/.style={fill=white},
xmajorgrids,
ymajorgrids,
legend style={legend cell align=left, align=left, draw=white!15!black},
title style={font=\Huge},xlabel style={font={\LARGE}},ylabel style={font=\LARGE},legend style={font=\large},
]
\addplot [color=blue, line width=2pt]
  table[x=t, y=e, col sep=comma]
  {Figures_tex/Data/Arbitrary_Drone_7.csv};
\addlegendentry{Drone 7}

\addplot [color=red, line width=2pt]
  table[x=t, y=e, col sep=comma]
  {Figures_tex/Data/Arbitrary_Drone_8.csv};
\addlegendentry{Drone 8}

\addplot [color=green!60!black, line width=2pt]
  table[x=t, y=e, col sep=comma]
  {Figures_tex/Data/Arbitrary_Drone_9.csv};
\addlegendentry{Drone 9}

\addplot [color=black, dotted, line width=1.0pt]
  table[row sep=crcr]{%
7.54166666666667	0\\
7.54166666666667	100\\
};
\addlegendentry{switch}

\addplot [color=black, dotted, line width=1.0pt, forget plot]
  table[row sep=crcr]{%
35.1666666666667	0\\
35.1666666666667	100\\
};
\addplot [color=black, dotted, line width=1.0pt, forget plot]
  table[row sep=crcr]{%
37.75	0\\
37.75	100\\
};
\addplot [color=black, dotted, line width=1.0pt, forget plot]
  table[row sep=crcr]{%
65.4166666666667	0\\
65.4166666666667	100\\
};
\addplot [color=black, dotted, line width=1.0pt, forget plot]
  table[row sep=crcr]{%
65.4583333333333	0\\
65.4583333333333	100\\
};
\addplot [color=black, dotted, line width=1.0pt, forget plot]
  table[row sep=crcr]{%
70	0\\
70	100\\
};
\addplot [color=black, dotted, line width=1.0pt, forget plot]
  table[row sep=crcr]{%
74.25	0\\
74.25	100\\
};
\addplot [color=black, dotted, line width=1.0pt, forget plot]
  table[row sep=crcr]{%
79.7083333333333	0\\
79.7083333333333	100\\
};
\addplot [color=black, dotted, line width=1.0pt, forget plot]
  table[row sep=crcr]{%
86.0416666666667	0\\
86.0416666666667	100\\
};
\addplot [color=black, dotted, line width=1.0pt, forget plot]
  table[row sep=crcr]{%
89.7916666666667	0\\
89.7916666666667	100\\
};
\addplot [color=black, dotted, line width=1.0pt, forget plot]
  table[row sep=crcr]{%
99.9583333333333	0\\
99.9583333333333	100\\
};
\addplot [color=black, dotted, line width=1.0pt, forget plot]
  table[row sep=crcr]{%
121.875	0\\
121.875	100\\
};
\addplot [color=black, dotted, line width=1.0pt, forget plot]
  table[row sep=crcr]{%
148.416666666667	0\\
148.416666666667	100\\
};
\addplot [color=black, dotted, line width=1.0pt, forget plot]
  table[row sep=crcr]{%
164.541666666667	0\\
164.541666666667	100\\
};
\addplot [color=black, dotted, line width=1.0pt, forget plot]
  table[row sep=crcr]{%
177	0\\
177	100\\
};

\end{axis}
\end{tikzpicture}%
}
    \end{subfigure}
    \caption{System responses to \texorpdfstring{$\cL_2$}{L2} disturbances: The dotted lines indicate the switching instants. The right plot presents magnified views of the corresponding left plot.} \label{fig:stability_analysis_result}
\end{figure}

The simulation is conducted with a sampling frequency of 24 Hz, corresponding to a sampling period of 41.7 ms, and a total duration of 180 s.
Switching occurs randomly 15 times over the entire simulation horizon, and only $i_0$ changes.
The results are presented in \autoref{fig:stability_analysis_result}, where the state error values are defined as $\sqrt{\sum_{p\in\bbN_9}\bx_p^T(t)\bx_p(t)}$.
The result shows that the switched system remains stable under arbitrary switching with disturbances.

The $\cL_2$ stability under arbitrary switching of the switched networked system can also be analyzed by considering the closed-loop dynamics of the system following
\begin{align*}
    \dbx=(\bA-\bB\bK\bH_{\sigma(t)}^c)\bx+\begin{bmatrix}
        \bB & -\bB\bK
    \end{bmatrix}\begin{bmatrix}
        \bu_p \\ \bu_c
    \end{bmatrix},\,
    \by=\bx
\end{align*}
where $\bA{=}\diag(\bA_p)_{p{\in}\bbN_9}{\in}\bbR^{108{\times}108}$, $\bB{=}\diag(\bB_p)_{p{\in}\bbN_9}\in\bbR^{108\times36}$, $\bK{=}\diag(\bK_p)_{p{\in}\bbN_9}{\in}\bbR^{36{\times}108}$, and $\bu_u{\in}\bbR^{36}$ and $\bu_c{\in}\bbR^{108}$ are the exogenous inputs of \acp{uav} and their controllers, respectively.
One approach is to search a common storage function and a common supply rate of the closed-loop switched networked system.
This can be implemented using two types of dissipativity variables: a full variable matrix, or a block-diagonal matrix of smaller variables, i.e., $\bX=\diag(\bX_p)_{p\in\bbN_9}$.
The former provides greater flexibility but requires significantly more computation time than the latter.

Second, the $\cL_2$ stability under arbitrary switching can be determined by employing the scattering function of the closed-loop switched system \citep{polushin2021stabilization}.
According to \cite[Thm. 1]{polushin2021stabilization}, the scattering function is obtained by analyzing the eigenvalues and eigenvectors of the active supply rates of each switching mode.
To apply this method, the liveness condition must be satisfied, and it is straightforward to verify that the considered system meets this requirement. 
The existence of this scattering function under this liveness condition ensures that the transformed switched dissipative system is $\cL_2$ stable under arbitrary switching.
The main advantage of this approach is that it explicitly provides a scattering function that modifies the system’s input and output to achieve the desired $\cL_2$ gain. 
However, this method requires the dissipativity parameters of each mode to be identified in advance, which is computationally demanding and may lead to conservatism in the analysis.


\begin{table}
\begin{center}
\caption{Computation time comparison}\label{tb:comp_time}
\begin{tabular}{L{2.9cm} L{2.7cm} C{1.25cm}}
\multicolumn{2}{c}{Approaches} & Time[s]  \\\hline
\multicolumn{2}{l}{\autoref{thm:stability_network_nonswitched_agents}} & 1.8496 \\
\multicolumn{2}{l}{Closed-loop analysis (diagonal variable)} & 3.0862 \\
\multicolumn{2}{l}{Closed-loop analysis (full variable)} & 264.0485  \\
\multirow{2}{*}{\cite[Thm. 1]{polushin2021stabilization}}& (eigenvalue analysis) & 0.0153 \\
 & (find parameters) & 2.2962 \\ \hline
\end{tabular}
\end{center}
\end{table}

\autoref{tb:comp_time} reports the computation time required to verify the $\cL_2$ stability under arbitrary switching of \autoref{eq:uav_dynamics}.
The closed-loop analysis is more time-consuming than the approach based on \autoref{thm:stability_network_nonswitched_agents}, even when diagonal variables are used, since it involves solving a large-scale \ac{lmi}.
The method in \cite[Thm. 1]{polushin2021stabilization} requires eigenvalue analysis of the given dissipativity parameters, which is impractical for optimal control frameworks.
In addition, although solving the eigenvalue problem to obtain the scattering function is not computationally expensive, applying \cite[Thm. 1]{polushin2021stabilization} still necessitates determining the dissipativity parameters of each switching mode in advance, which incurs a computational cost comparable to that of the closed-loop analysis. 
By contrast, \autoref{thm:stability_network_nonswitched_agents} relies only on smaller \acp{lmi} whose sizes match the dimension of the individual \ac{uav}.
This substantially reduces computation time and enables distributed analysis of the switched networked system.

\section{Conclusion}
This paper has established a framework for analyzing the switched dissipativity and $\cL_2$ stability of switched and large-scale networked systems.
The notion of $\cL_2$ stability with dissipativity is extended to switched systems by introducing the concept of $\cL_2$ stability under arbitrary switching.
A switched networked system is considered where the network topology switches with dissipative agents. 
In this case, $\cL_2$ stability under arbitrary switching can be verified by analyzing the open-loop dissipativity of each agent, which generalizes the \ac{vndt} in \citep{moylan1978stability,vidyasagar1981input}.
A numerical example shows that the proposed switched network dissipativity theorem validates $\cL_2$ stability under switching network topologies through open-loop analysis of individual agents, offering a more computationally effective alternative to other dissipativity-based and Lyapunov-based methods.
The proposed works are expected to be valuable for distributed stability analysis or controller synthesis of switched networked systems.




\bibliography{MyBib}             

\end{document}